# Subjective Functionality and Comfort Prediction for Apartment Floor Plans and Its Application to Intuitive Searches

Taro Narahara, *Member, IEEE,* and Toshihiko Yamasaki, *Member, IEEE,*

*Abstract*—This study presents a new user experience in apartment searches using functionality and comfort as query items. This study has three technical contributions. First, we present a new dataset on the perceived functionality and comfort scores of residential floor plans using nine question statements about the level of comfort, openness, privacy, etc. Second, we propose an algorithm to predict the scores from the floor plan images. Lastly, we implement a new apartment search system and conduct a large-scale usability study using crowdsourcing. The experimental results show that our apartment search system can provide a better user experience. To the best of our knowledge, this study is the first work to propose a highly accurate prediction model for the subjective functionality and comfort of apartments using machine learning.

*Index Terms*—Real estate floor plans, crowdsourcing, graph analysis, attractiveness prediction

## I. Introduction

RECENTLY, the real estate industry has been showing increasing interest in applying machine learning-assisted tools, such as price prediction [1], [2], [3], [4] and apartment-searching [5] tools. Some online platforms can help users to search real estate properties by specifying metadata, such as type of apartment and room size. However, many users inspect floor plans based on more intuitive sensory impressions, such as feeling of *living comfort* and presence of *openness* and *privacy*. This makes it difficult to estimate the perceptive values of apartments through any currently available retrieval system, as there are no quantifiable data that represent such subjective characteristics of apartments. Moreover, apartment properties listed with the same size and type in their metadata (e.g., two-bedroom apartments) could feature different room arrangements, which will have a significant impact on their functionality and overall impressions. Further understanding the relationships between floor plan images and structured data, including the connectivity of rooms and metadata, could improve such search platforms and user experience.

Information on floor plans has been widely adopted by users to evaluate the values of properties over the years, as can be seen in many real estate portal sites today. A floor plan image of an apartment includes various room types, room sizes, and connections and spatial layouts of the rooms. In a customer survey conducted in Japan[1], a floor plan is listed in the top five priorities for customers for their apartment searches, and customers are very reluctant to compromise on their preferred floor plans and are often willing to invest more for their pursuits. Essential elements that largely influence functional, environmental, and some perceptive characteristics of apartments, such as locations of walls, columns, windows, and wet areas, are already set in floor plans, no matter how we change final finish materials and furniture. Although we can estimate the subjective quality of apartments, such as their *living comfort*, simply by looking at the floor plan images, no related work nor dataset has been reported for such a task.

In this study, therefore, we first construct a new dataset that contains 1,000 floor plan images (Hereafter, referred as dataset A). Each image has a subjective score from nine perspectives relating to perceived quality and functionality of the apartments. Examples of the generated dataset are shown in Fig. 1. Based on this dataset, we present a multimodal neural network-based framework to predict subjective apartment scores via their floor plan images, graph representations, and various metadata. The experimental results show that we can predict the scores with a correlation coefficient of 0.701 on average. This is relatively high, considering that they are all subjective scores.

The contributions of this paper are summarized as follows.
- We created and analyzed a large-scale dataset of subjective evaluations of both perceived quality and functionality of real estate floor plan images using crowdsourcing.
- Our proposed prediction model that extracts features from floor plan images and their graph structures proved to be effective and highly accurate.
- Using a new set of floor plan images with predicted scores by our model as a database, our new apartment search tool that can query the functionality and comfort can provide a better user experience based on our user study.

## II. Related Works

### A. Real Estate Tasks using Property Images

Several studies have worked on tasks related to real estate using property images. In [2], they improved the accuracy of real estate price prediction by predicting the luxury levels of the rooms using the appearance and interior images of properties. Law et al. [6] showed that the street view and satellite images are also helpful when predicting house prices. In [7], the researchers attempted to predict the construction age of the property by combining the predictions for each salient image patch of the property, resulting in greater accuracy than human predictions.

[1]https://suumo.jp/article/oyakudachi/oyaku/chintai/fr_data/hikkoshi-sumikae2017/, accessed 09/16/2020

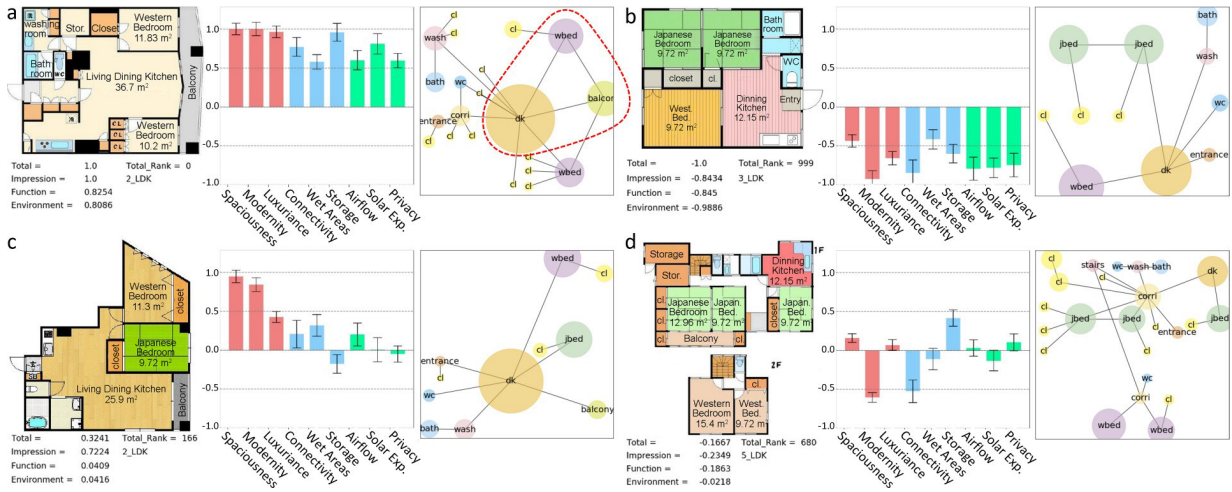

Fig. 1. Examples of our generated functionality evaluation dataset using real estate floor plan images. Each example shows a floor plan image, a bar graph for nine evaluation measures, and a graph from left to right. (a) Highest scoring floor plan; (b) lowest scoring floor plan; (c) floor plan that scored high in the modernity; (d) floor plan that scored low on modernity.

Moreover, deep learning has been applied to real estate property images, and some studies have analyzed real estate images themselves. In [8], real estate images were classified into different types (e.g., bedroom, kitchen, living, and garden) by employing contrast-limited adaptive histogram equalization (CLAHE) and applying long short-term memory (LSTM) in both vertical and horizontal directions. Wang et al. [9] predicted which of two images of the same property is more attractive by a pairwise comparison network.

These works demonstrated the use of property images in specific tasks that are related to the appearance of the property. In our task, to comprehensively represent the user experience of the perceived quality and functionality, we assume that the floor plan information with related metadata can provide more structure and detailed information on the property.

*B. Real Estate Tasks using Floor Plan*

Several related works have been conducted on floor plan image analysis. Before the development of deep learning, some researchers manually and graphically analyzed floor plans using adjacency graphs (with rooms, corridors, and other features as labeled nodes) and used them to classify apartments into different types [10]. Takizawa et al. [11] analyzed the relationships among adjacency graph structures of apartment floor plans and their rental fees. They extracted subgraphs from the adjacency graphs and effectively estimated the rent from the presence and absence of common subgraphs. However, the cost of creating adjacency graphs of floor plans by hand was very high.

Floor plan images are also proved to be useful for rent price prediction [12], [13], [14]. In [12], [13], it has been demonstrated that conventional bag-of-features (BoFs) [15] has a potential to achieve lower-error prediction with smaller variance though the contribution of the BoFs is smaller than the other apartment attribute information. In [14], image features were applied to hedonic price models [16] to predict the apartment rent price after controlling for locational and structural characteristics of an apartment.

Recently, machine learning has been applied to the analysis of floor plan images. Yamasaki et al. [5] used deep neural networks (DNNs) to conduct semantic segmentation of floor plan images. They further developed the method to systematically generate adjacency graphs of floor plans from the semantically segmented images.

Takada et al. [17] utilized multi-task learning for floor plan images and retrieved similar floor plans to the query. The floor plan recognition was then applied to property recommendation [18], [19] and retrieval [20]. Additionally, a toolbox for converting floor plan images to a vector format is presented in [21]. Furthermore, generating floor plans using graphs [22], [23], panoramic images [24], or 3D scans [25], [26], [27], [28], [29], [30] is emerging. Generating furniture layouts using graphs is also discussed in [31].

To the best of our knowledge, this study is the first to propose an accurate prediction model for subjective scores of apartments using machine learning.

*C. Prediction of Subjective Scores of Multimedia Content*

Assessing the perceived low-level quality of images [32], [33], [34], [35] and videos [36], [37], [38], [39] has been an important topic in multimedia. These works tried to predict the perceived quality when quality of the multimedia content is somewhat degraded by low-level factors such as compression and noise.

Predicting higher-level subjective scores has also been a very active research area. For instance, image aesthetics analysis care more about the color usage, composition, etc., not the low-level noise in the content [40], [41], [42], [43], [44], [45].

Sentiment and emotion classification and affective analysis another direction of research for analysing subjective evaluation of multimedia content. A lot of related works can be found for texts [46], [47], [48], [49], images [50], [51], [52], [53],

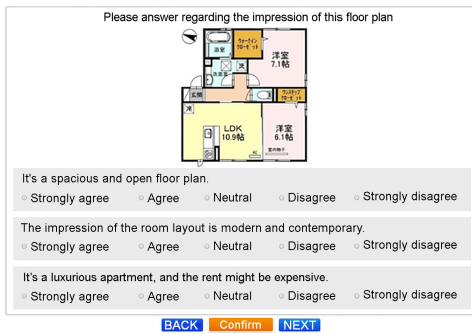

Fig. 2. A screenshot of the crowdsourcing page.

[54], [55], [56], [57], speech [58], music [59], videos [60], [61], [62], [51], [63], and their combinations [64], [65], [66].

Skills and creativeness have also been research targets such as skill assessment [67], [68], creativity [69], click through rate prediction for online advertisements [70], [71], presentation slide assessment [72], [73], and so on.

Subjective evaluation for apartment floor plans is the first work to the best of our knowledge. In the paper, we show that predicting the subjective functionality and comfort is possible, and we also demonstrate possible applications of such prediction.

## III. DATASET CREATION

### A. Subjective Scores

In this study, we used crowdsourcing to create a new dataset based on subjective evaluation of real estate floor plan images through a set of statements that question their levels of comfort, openness, privacy, and other characteristics. In total, 3,128 workers participated in this evaluation. We recruited 400 participants separately from 10 groups: two genders X five age ranges (20-29, 30-39, 40-49, 50-59, and 60+) using a crowdsourcing service. We did not get all responses and also removed those who chose the same rating for all (i.e., "straight-lining") and obtained 3,128 participants' results.

We used floor plan images of Japanese rental apartments from the "Home's dataset" released by LIFULL Co., Ltd. with the cooperation of the National Institute of Informatics[2], which has been widely used as a general floor plan image dataset in our international research community [74], [75], [22], [76]. We randomly selected 1,000 floor plan images that included apartments with one, two, three, and four or more bedrooms in balanced proportions and prepared the following nine question statements for each image (questions Q1 to Q3 are about impressions, Q4 to Q6 are about functionality, and Q7 to Q9 are about environmental criteria):

- Q1 (Spaciousness): It is a spacious and open floor plan.
- Q2 (Modernity): The impression of the room layout is modern and contemporary.
- Q3 (Luxurance): It is a luxurious apartment, and the rent might be expensive.

[2]National Institute of Informatics (NII), https://www.nii.ac.jp/dsc/idr/lifull/homes.html (accessed: 05.05.2020)

- Q4 (Connectivity): Connectivity, adjacency, and layout of rooms and circulation for residents are efficient and look comfortable.
- Q5 (Wet Areas): The traffic paths for the kitchen, bathroom, and restroom are good.
- Q6 (Storage): Locations and sizes of storage spaces are good.
- Q7 (Airflow): Airflow inside the floor plan is good overall.
- Q8 (Solar Exp.): Solar exposure inside the floor plan is good.
- Q9 (Privacy): The arrangement and adjacency of rooms fully consider the privacy of each family member.

In Q9, participants were provided a family size to evaluate the privacy of the image. Each floor plan was evaluated with a five-grade score on a scale of 1 (strongly disagree) to 5 (strongly agree) by participants. Only the floor plan images were shown to participants for the evaluation (Fig. 2). Each participant was asked to evaluate 25 floor plan images. As a result, each floor plan was evaluated by 78 to 80 individuals (i.e., 2 genders × 5 age ranges × 400 participants × 25 images / 1,000 total images = 100 participants per image, about 20% dropped out, which coincide with 3128/4000 = 0.782). Each participant would have different mean and standard deviation in their evaluation scores, and therefore the scores were standardized before taking the average among the participants.

We used the value obtained by subtracting the mean value from each raw score and dividing it by the standard deviation. The resulting scores were normalized between -1 and 1 for each question. Figure 1 shows selected examples of floor plans in dataset with the corresponding results of the nine scores of subjective measure. The error bar represents the standard deviation.

Next, for each floor plan, we averaged the scores from the nine questions. Then, we standardized those mean values from all floor plans and normalized them between -1 and 1, and called the value the "Total Score" of the floor plan.

### B. Segmentation and Graph Extraction

A floor plan can also be viewed as a graph with nodes representing rooms and with edges representing connections between them [5], [22], [23], [20]. Therefore, we extracted graphs from floor plan images and used them as features to predict the nine subjective scores defined above. This is a reasonable assumption because floor plan images are inspected for the connections among rooms and their adjacency.

In order to extract corresponding graphs automatically from floor plan images, we used the following two steps (Fig. 3 top left). First, we prepared 3,800 new floor plan images (no overlap with the dataset A), which we call dataset B, with their manually annotated segmented images using an online annotation tool (please see Fig. 3). They were consistently color-coded and semantically segmented into the following 15 classes of elements: wall, western bedroom (wbed), Japanese bedroom (jbed), dining kitchen (dk), restroom (wc), bathroom (bath), washroom (wash), balcony (balc), entrance (ent), corridor (corri), stairs, closet (cl), door, window, and unknown elements that do not belong to any category (abbreviations in parentheses are used in figures representing graph nodes). Then we prepared a segmentation prediction network based on the method introduced in [5] using an improved network architecture, DeepLab [77], instead of the fully convolutional networks. Using 3,040 for training, 380 for validation, and 380 for testing, we trained the network using the segmented images as ground truth data. We automatically obtained 1,000 segmented images with subjective scores for our dataset by feeding the dataset A into this pre-trained network using DeepLab.

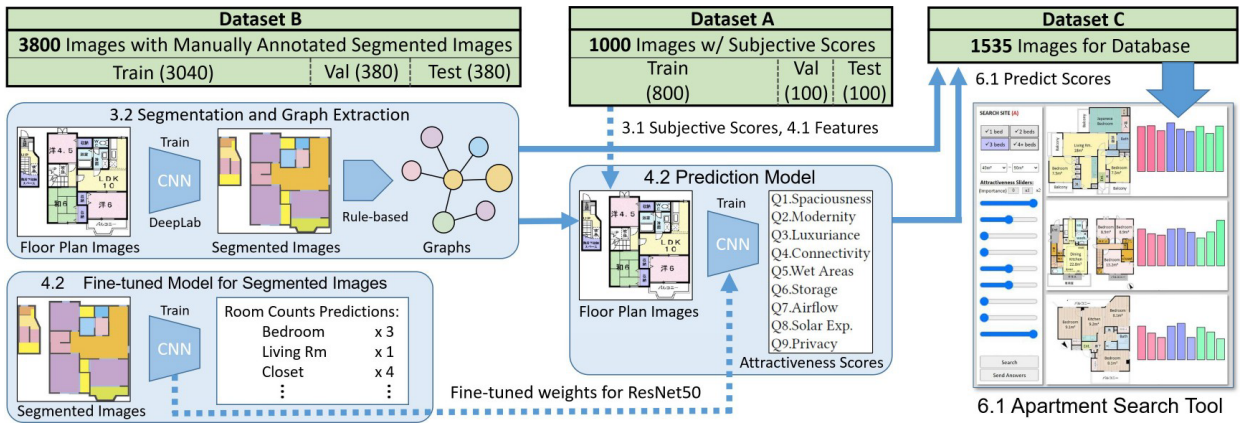

Fig. 3. Overview of our framework

Second, we used the rule-based method to extract 1,000 graphs from the segmented images following the same procedure in [20]. We used 11 elements for nodes of the graphs, excluding the wall, door, window, and unknown elements from the above 15 elements. We added edges between two rooms only if they are directly accessible through a physical opening or door. The resulting dataset was used to determine whether the differences in graph structures influence the impression and functionality of apartments from subjective evaluations.

### C. Overview of Our Datasets

Figure 3 shows an overview of our framework. To predict subjective scores only from the real-estate floor plan images, we prepared three sets of floor plan images without any overlaps to avoid bias for network models' training. The first set of 1,000 images (dataset A) was used for the dataset with subjective scores and to train our prediction model in Section IV. The second set with 3,800 images (dataset B) is used to obtain the network that outputs segmented images from the floor plan images as input. We also used the dataset B to train an ImageNet-pretrained feature extractor network using the color-coded semantically segmented images as shown in Fig. 6. Finally, we predicted subjective scores by our model from a separate set of 1,535 floor plan images (hereafter, we will call this dataset C and please see Fig. 3), and it is also used for our proposed apartment search tool in Section VI.

## IV. Proposed Methods

In this section, we introduce our prediction model. We propose to use two types of inputs for the model: floor plan images and structured data features. For the images, we used both floor plan images and semantically segmented images from Section III. In Section IV-A, we explain our methods to extract four features from the structured data based on graphs and metadata of floor plans. The image features are described in Section IV-B. In Section IV-C, we introduce our prediction model in detail.

### A. Features

*1) Emerging Subgraphs:* To extract frequently appearing common subgraphs that are very important and contribute to either higher or lower evaluation scores for each question for the floor plans, we first performed frequent subgraph mining on a total of 1,000 graphs in the dataset A using graph-based substructure pattern mining (gSpan) [78]. For the condition to evaluate subgraph isomorphism, we considered node attributes that represent the room types for graph matching. The edge attributes that represent door and window types were not considered. As a result, 162,470 subgraphs were extracted by setting the minimum support threshold to 5 (i.e., the condition for extracting the common subgraph corresponding to at least 5 out of 1,000). The following three steps were further performed to extract subgraphs that are more relevant to each subjective score.

*Step* 1: For each question from Q1 to Q9 and the "Total Score" from Section III-A (10 items in total), common subgraphs that were included in the floor plan graphs with evaluation scores of the top and bottom 10% were extracted and separated into a total of 20 classes. We further narrowed down the selections of subgraphs by setting the following three thresholding strategies. First, the minimum support threshold was set to 10. Second, the average score of the apartments that contain the target subgraph should be 0.25 or larger for those in the top 10% classes and -0.25 or lower for those in the bottom 10% classes. Third, we also used the growth rate (GR), which is widely used for discovering discriminative patterns in emerging pattern mining [79]. The ratio between the GR for the top 10% and that for the bottom 10% should be larger than 4 or smaller than 1/4. These thresholds were set by our empirical study.

*Step* 2: To further extract relevant common subgraphs in the above 20 classes, we considered identification numbers of extracted subgraphs as "words" in 20 different "documents." Then we performed term frequency - inverse document frequency (TF-IDF), which evaluates how relevant a word (subgraph) is to a document (class) in a collection of documents (classes that represent each question). Based on the obtained TF-IDF weights, we sorted important subgraphs that were relevant to the top or bottom 10% of floor plans evaluated based on a specific question in each class. The use of TF-IDF helped eliminate frequently appearing subgraphs found in multiple classes not yet relevant to any specific classes.

For any subgraph that included other subgraphs (i.e., inclusion dependency) within the same class, subgraphs with a minimum number of nodes were always kept. If the larger-size subgraph in the top 10% class has the larger mean evaluation score than those of the minimum-size subgraphs defined above, it is also kept (and vice versa for the bottom 10% classes).

*Step* 3: We sorted all the extracted subgraphs from the previous step based on the mean evaluation score in each class and obtained the top 20 subgraphs for each of the 20 classes. By eliminating duplicates that appeared in multiple classes from a total of 400 extracted emerging subgraphs, we obtained 230 unique emerging subgraphs. For each floor plan in the dataset, a 230-dimensional feature vector that contained 0 or 1 based on the absence or presence of 230 emerging subgraphs, respectively, were extracted.

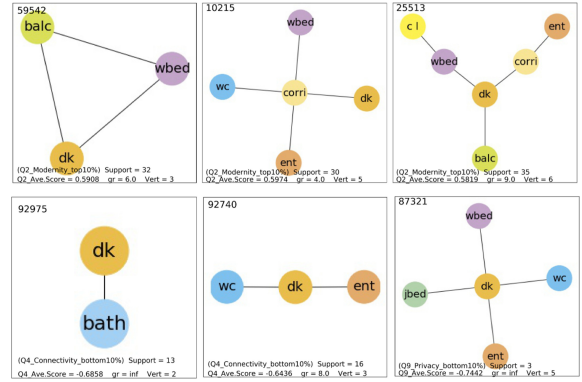

Fig. 4. Examples of emerging subgraphs frequently found in floor plans having a high Q2 (modernity) score (top row). The top left also appears in Fig. 1a (circled in red). Examples found in floor plans having a low Q4 (connectivity) (bottom left & middle) and a Q9 (privacy) score (bottom right).

Figure 4 shows examples of the emerging subgraphs that were extracted using the above three steps. The subgraph on the top left is a triangle graph connecting (wbed)-(dk)-(balc) as a subgraph that appears in the top 10% of the Q2 (modernity) class. For example, the floor plan in Fig. 1a includes this subgraph and has a very high evaluation score for Q2. It represents the wide balcony space that is open to both the bedroom and kitchen and is often considered a contemporary layout among Japanese apartments advertised as a "wide-span balcony". The subgraph with two nodes, (dk)–(bath), was in the class representing the bottom 10% of the Q4 (adjacency and connectivity) class (bottom left in Fig. 4). This has a bathroom door that immediately opens to a living room, allowing anyone to enter into a communal space directly after taking a bath, which is not considered desirable. Adding a washing room between the two nodes is a common practice in Japan, as it provides a room to change clothes in before entering the bathroom. This subgraph was also included in the class for the bottom 10% for the Q9 (privacy) class. Our method was able to extract emerging subgraphs representing such characteristics of floor plans in 20 classes.

*2) Subgraphs for Wet Areas:* Wet areas are particularly important because they are more personal spaces. Therefore, we paid special attention to them. For each of 1,000 floor plans in the dataset A, a connected subgraph that contained rooms related to the usage of water, including dining kitchen (dk), washroom (wash), bathroom (bath), and restroom (wc), were extracted. Any nodes bridging the above four nodes, such as a corridor (corri), that were necessary to form a single connected subgraph that contained all four of the above nodes, were also included depending on the floor plan layouts. If there were more than two such bridging nodes and only one node was sufficient to form a connected subgraph, the node that served a communal use or as circulation, such as corridor, was selected over a node use for a private purposes, such as a bedroom. In total, 162 unique subgraph types for water-related rooms were extracted. Out of all the floor plans, 76% belonged to the five types. 120 types of subgraph appeared only once and therefore they are discarded. As a result, 42 subgraph types are used in this study. Two out of the five remaining types of subgraphs included three nodes that were directly linked: (wc)–(wash)–(bath). However, this is not a desirable layout, as its circulation paths for (wc) and (bath) are crisscrossed at the washing room (Fig. 5). It forces one to enter into (wash) while another is still present after using either room. A mean evaluation score for Q5 (wet areas) for those two types were lower (0.171 and 0.208) than the mean scores from the other three types (0.330, 0.296, and 0.249). As a result, we extracted a 42-dimensional one-hot feature vector based on the presence of the 42 types of subgraphs for wet areas.

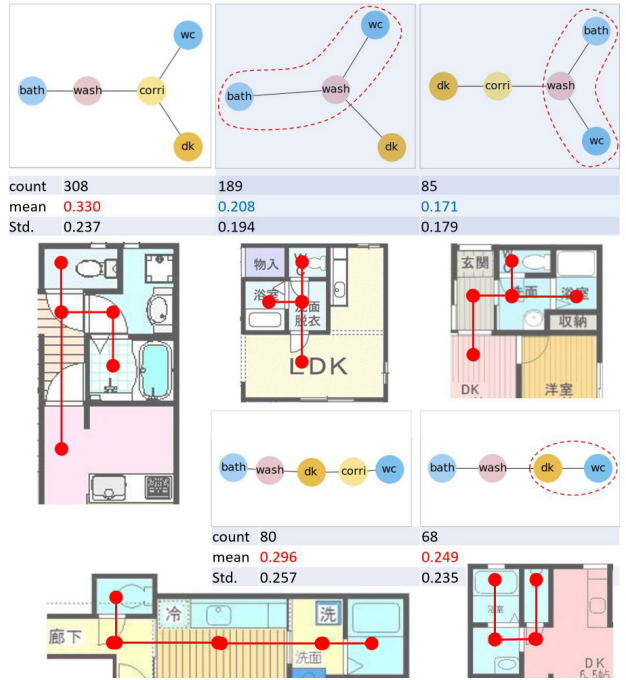

Fig. 5. The main five types of subgraphs for wet areas that make up 76% of all floor plans. Two subgraphs that include the linkage (wc)–(wash)–(bath), circled in red, have lower mean scores for Q5 (wet areas) than the others.

*3) Feature Based on MCS Graph Similarity:* The similarity between graph structures of all 800 floor plans in the training set of the dataset A was calculated based on the method of [80], [17], [81] using the maximum common subgraph (MCS) as a graph similarity measure. The similarity was 1 when the two graphs are perfectly matched, and 0 when there were no common parts. Any input graph could be expressed by a 800-dimensional vector that represents distances from each of the 800 graphs in the training set of the dataset A. We used this vector based on the MCS similarity to extract a feature that represented an entire (global) characteristic of each apartment, as opposed to a local sub-structure from a subgraph.

*4) Feature Based on Metadata:* In addition to the above three features extracted from graph structures, we listed areas and numbers of room types using the metadata for each floor plan. This resulted in 30-dimensional feature vectors that represented the areas and numbers of room types.

From the above, we obtained four feature vectors that represented structural data of floor plans based on emergent subgraphs, subgraphs for wet areas, MCS graph similarity, and metadata. All feature vectors were standardized before using them for machine learning (described in the next section) such that their distributions had a mean value of 0 and a standard deviation of 1.

### B. Image Features

We also fed the network two types of images: floor plan images and consistently color-coded semantically segmented images (prepared for the automated graph generation in Section III). The inputs to the network were two sets of RGB images of resolution 224 × 224.

First, for the floor plan images, using ResNet50 [82] pre-trained on ImageNet [83], a 2,048-dimensional vector of the pool5 layer was extracted from deep features of the images. Then a new FC layer was added in place of the original one.

Second, for the segmented images, we used the same network architecture, but the network was pre-trained first using ImageNet and then fine-tuned for the multi-task classification task to predict the number of the rooms of the 13 room types excluding the wall and unknown elements from the 15 types defined in Section III-B. Similarly, a 2,048-dimensional vector of the pool5 layer was extracted.

Two features extracted from the two sets of images described above were also added the same way as above, followed by the FC, batch normalization (BN) [84], and dropout [85] layers. Finally, these two added features from the structured data and images were added again, followed by the FC and dropout layers. The final FC layer generated a value predicting the score (see Fig. 6 for details). For the FC layers, we used Leaky Relu as an activation function.

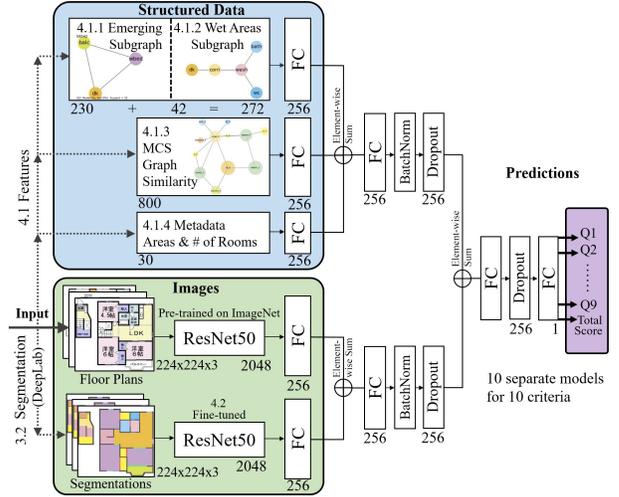

Fig. 6. Proposed network architecture.

### C. Prediction Model

Figure 6 shows our proposed network architecture to predict the evaluation scores from Q1 to Q9 and the "Total Score" (Ten separate models in total).

Two features based on emergent subgraphs (IV-A1) and subgraphs for wet areas (IV-A2) were concatenated into one 272-dimensional vector, which was reduced to 256 dimensions using a fully connected (FC) layer. The 800-dimensional feature vector from the MCS graph similarity (IV-A3) and the 30-dimensional feature based on metadata, such as areas and numbers of room types (IV-A4), were also reduced to 256 dimensions using FC layers. Then, the above three input features were added, followed by the FC, BN, and dropout layers. The BN and dropout layers were added to improve generalization and to reduce overfitting.

## V. EXPERIMENTS

We used our dataset created in Section III. We divided the dataset into 800 floor plans for training, 100 for validation, and 100 for testing. We trained 10 separate models for predicting 10 scores using the network in Section IV-C because it was better than multi-task learning using a single model in our preliminary experiment. Regression models were created for each using the mean squared error (MSE) as a loss function, and we trained these networks using 800 training images. We applied the Momentum Stocastic Gradient Descent (SGD) algorithm to train models with a batch size of 20, a learning rate of $10^{-3}$, a decay rate of $2.86 \times 10^{-5}$, and a momentum of 0.9 for the 35 epochs. The system was implemented using Keras[3]. We applied the Pearson correlation coefficient (PCC) to measure the correlation between the predicted values and the evaluation scores from the dataset as ground truths for 100 floor plans in the test data. We repeated the above process five times using a different randomly selected set of 800 training, 100 validation, and 100 test floor plans each time, and used the mean values of all five results as the final

---
[3]https://keras.io

TABLE I
PREDICTION ACCURACY COMPARISON. VALUES ARE PEARSON CORRELATION COEFFICIENTS (PCC). BLACK CELLS REPRESENT PCC>0.7 AND GRAY CELLS INDICATE PCC<0.7.

| Model Features w/o | w/o SubG | w/o MCS | w/o Meta | w/o Img | w/o Segm | IV-C Prop. |
|---|---|---|---|---|---|---|
| Q1. Spaciousness | 0.669 | 0.730 | 0.707 | 0.637 | 0.648 | 0.721 |
| Q2. Modernity | 0.782 | 0.787 | 0.772 | 0.771 | 0.765 | 0.793 |
| Q3. Luxuriance | 0.762 | 0.753 | 0.751 | 0.749 | 0.747 | 0.776 |
| Q4. Connectivity | 0.542 | 0.620 | 0.592 | 0.546 | 0.571 | 0.637 |
| Q5. Wet Areas | 0.477 | 0.452 | 0.422 | 0.439 | 0.499 | 0.525 |
| Q6. Storage | 0.715 | 0.751 | 0.728 | 0.713 | 0.733 | 0.751 |
| Q7. Airflow | 0.494 | 0.514 | 0.573 | 0.446 | 0.592 | 0.607 |
| Q8. Solar Exp. | 0.520 | 0.571 | 0.531 | 0.525 | 0.564 | 0.591 |
| Q9. Privacy | 0.764 | 0.780 | 0.786 | 0.791 | 0.748 | 0.816 |
| Total Score | 0.727 | 0.740 | 0.763 | 0.712 | 0.755 | 0.794 |
| Average | 0.645 | 0.670 | 0.662 | 0.633 | 0.662 | 0.701 |

PCCs for the proposed as well as the following comparative baseline models, which were prepared to compare and verify the prediction results.

*A. Baseline Methods*

As baseline methods, we prepared the following five models using DNNs by subtracting one of five input features from the proposed network in Fig. 6:

*w/o SubG:* The model without the input features based on emergent subgraphs (IV-A1) and subgraphs for wet areas (IV-A2).
*w/o MCS:* The model without the input features based on the MCS graph similarity (IV-A3).
*w/o Meta:* The model without the input features based on metadata such as areas and numbers of room types (IV-A4).
*w/o Img:* The model without the features based on floor plan images.
*w/o Segm:* The model without the input features based on consistently color-coded semantically segmented images.

*B. Results*

First, we discuss the segmentation accuracy using the dataset B. The mean IoU for the test data was 82.1%, and the average mean accuracy was 89.0% for the test set. Although the segmentation accuracy is not perfect, this is acceptable considering the prediction accuracy of functionality and comfort prediction as discussed below.

Table I shows the results of both the proposed and baseline models using the test set of the dataset A. Our proposed model outperformed all of the comparative baseline models in terms of the mean value of PCCs from all 10 criteria (0.701) and recorded the highest PCC for 9 out of 10 criteria (Table I). Therefore, it can be inferred that the network compensates for the weakness of each feature by combining them. The Pearson correlation coefficients (PCC) for our proposed model were over 0.7 for five criteria (Q1=0.721, Q2=0.793, Q3=0.776, Q6=0.751, and Total Score=0.794), and over 0.8 for the "privacy" criteria (Q9=0.816). In conclusion, although the measurement of the functionality of floor plans is a very subjective problem, our proposed prediction models are able to achieve a very strong correlation with human evaluation.

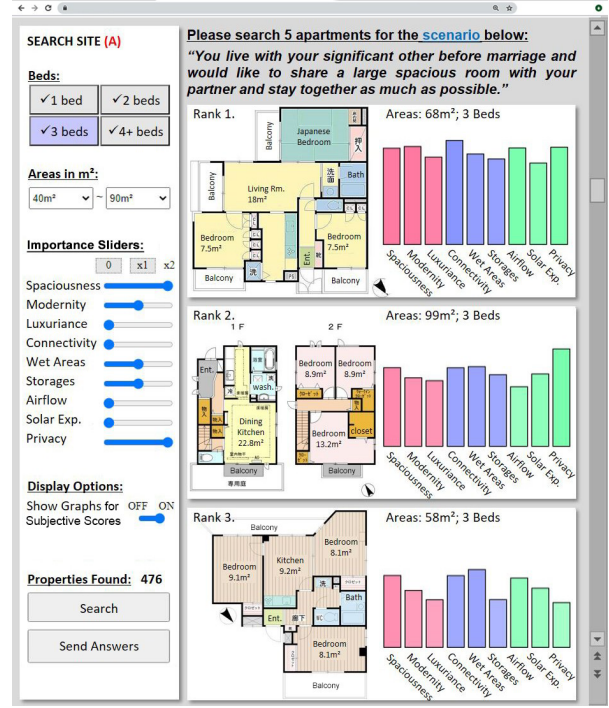

Fig. 7. The proposed apartment search tool

## VI. USABILITY STUDY

*A. Implementation*

We introduce a new interface for an apartment search tool that implements functionality and comfort as query items. In addition to a common search interface based on user selection of the number of bedrooms and a range of areas, our tool offers *importance sliders* with adjustable weights for importance in three levels: 0 (none); ×1 (important); and times ×2 (very important), for nine subjective criteria (see Fig. 7). After a user presses the search button, the tool calculates the scores for all floor plans based on the weighted sum of the predicted subjective values using the information from the sliders, and displays the scores in ranked order. This feature allows a user to search apartments based on a controlled weighted priority for qualitative criteria (i.e., extremely spacious apartments with sufficient privacy and storage spaces).

For our proposed search tool's floor plan database, we prepared a new set of 1,535 floor plan images that included apartments with one, two, three, and four or more bedrooms in balanced proportions (dataset C). These images were completely unbiased and were not used in any previous section in this paper. From the pre-trained network using DeepLab and the rule-based method in Section III-B, we obtained the 1,535 corresponding segmented images and graphs. Next, we executed the same procedures in Sections IV-A and IV-B to extract the five features from these images, which were then used to obtain the predicted attractiveness scores for the nine criteria for each of the 1,535 floor plans using the best model (i.e., the one with the highest PCCs) from the proposed network introduced in Section IV-C.

## B. Procedure

We evaluated our method through a large-scale user study. Among 200 participants recruited through a crowdsourcing service, we removed three who chose the same rating for all questions (i.e., "straight-lining"), removed six who did not complete the survey, and obtained 191 with a wide variety of attributes. Eighty-one participants identified themselves as male and 109 as female. The numbers of participants in their 20s, 30s, 40s, 50s, and over 60 were 30, 78, 53, 20, and 4, respectively. Thirty-eight currently live alone, and 81 participants were identified as married. Those living in urban areas numbered 115, while 60 lived in suburban areas and 11 in rural areas.

To make the study task realistic and providing internal motivation, we prepared hypothetical scenarios with different demands from five unique families and asked participants to search five apartments for each scenario. We set scenarios based on five completely different family structures to avoid demographic background bias of the participants.

- "You are a married couple, and your two children need their own rooms soon. You want a functional floor plan layout and don't want to pay extra for unnecessarily large spaces."
- "You are a family of five, living with one child and your spouse's parents. A well-functioning home with large storage spaces, kitchen, and wet areas are your top priority since you have a big family."
- "Due to the COVID-19 pandemic, you have been working from home and would like to have your own study room. You and your spouse want to have separate rooms to respect each other's privacy."
- "You live with your significant other before marriage and would like to share a large spacious room with your partner and stay together as much as possible."
- "You and your partner are a young couple and hope to have at least one child in the future. Ideally, you prefer to have a large enough balcony for your family to spend the weekend together."

We assigned these tasks for both our proposed tool and another baseline tool. The baseline tool represented commonly available real estate portal sites without our proposed features, featuring a search interface for a user to select the number of bedrooms and a range of areas. We studied major real estate portal sites [4,5,6,7], and found the above two items as common search features. The differences between the proposed and baseline tool are that the baseline tool has an identical interface except that it does not have functionality and comfort related options. As we wanted participants to focus on analyzing information only readable from floor plans to enable the fair comparison to our proposed tool, we excluded other common search features based on the location and cost of properties. The study was a within-participants design in which participants used both tools (counterbalanced across participants) and provided feedback on them.

[4] https://lifull.com
[5] https://www.livable.co.jp
[6] https://suumo.jp
[7] https://www.redfin.com

TABLE II
THE DIRECT COMPARISON QUESTIONS WERE ASKED ON A 5-POINT LIKERT SCALE. A HIGHER SCORE INDICATED A PREFERENCE FOR OUR PROPOSED TOOL, WHILE A LOWER SCORE INDICATED A PREFERENCE FOR THE BASELINE TOOL. A SCORE OF 3 INDICATED NO PREFERENCE.

| Question | Mean | CI | p-value |
| --- | --- | --- | --- |
| Which tool helped you find more desirable floor plans? | 4.03* | [3.90, 4.17] | $4.9 \times 10^{-34}$ <0.001 |
| Which tool did you enjoy better while searching? | 3.97* | [3.83, 4.11] | $1.6 \times 10^{-30}$ <0.001 |
| Which tool was faster for you to search floor plans? | 3.84* | [3.67, 4.01] | $1.1 \times 10^{-18}$ <0.001 |
| Which tool was easier to search apartments? | 3.76* | [3.60, 3.93] | $2.5 \times 10^{-16}$ <0.001 |
| Which tool was more intuitive for you to search floor plans? | 3.43* | [3.25, 3.62] | $7.0 \times 10^{-6}$ <0.001 |

* Significantly different based on 95% confidence interval.

TABLE III
THE FIVE-GRADE RATINGS OF ALL FLOOR PLANS SELECTED BY PARTICIPANTS USING TWO METHODS

| | Baseline | Proposed | p-value |
| --- | --- | --- | --- |
| Mean scores of the 5-grade ratings of all floor plans selected | 3.75 | 4.01 | $1.67 \times 10^{-34}$ <0.001 |

A t-test was conducted on two related scores; the average scores of five floor plans for five scenarios from all participants using both tools.

Participants were asked to search five apartments that met needs from each of the five scenarios using one tool, and switch to the same searches using the other tool (with the order of tool counterbalanced across participants). After completing both searching tasks, participants completed a post-experiment survey about their experiences. The survey asked participants to directly compare their experience with the proposed tool to the baseline tool in five Likert scale questions (shown in Table II). We also asked them to rate 50 selected floor plans (i.e., five plans × five scenarios × two tools) in a five-grade score on a scale of 1 (very unsatisfied) to 5 (very satisfied) (i.e., each participant gave a 1 to 5 score for each retrieved floor plan image.).

## C. Results

Overall, the participants showed a significant preference for our proposed tool for all five direct comparison questions in Table II. Participants appreciated the proposed tool. Compared to the baseline tool, it helped them find significantly more desirable floor plans (M = 4.03, p=$4.9 \times 10^{-34}$<0.001), and gave them a significantly more enjoyable experience (M = 3.97, p=$1.92 \times 10^{-30}$<0.001).

Despite the fact that the participants spent a longer average time to complete a search task using the proposed tool (M = 158.0s, Median=131s, SD = 105.7 s) than the baseline tool (M = 129.0s, Median = 105s, SD = 97.4s), the result from the direct comparison question shows that participants felt that they found the desired floor plans faster using the proposed tool (M = 3.84, p=$1.1 \times 10^{-18}$<0.001). The numbers of clicks on buttons per search increased using the proposed tool (M=18.3, Median=17, SD=7.6) compared to the baseline tool (M=13.2, Median=12, SD=5.9). These results show that our proposed tool required more time and mouse clicks for users due to additional features not in the baseline tool (e.g., importance sliders). However, these additional features did not lead them to think the proposed system is harder to use. Participants felt that the proposed tool made it significantly easier (M = 3.76, p=2.5×10-16<0.001) and more intuitive (M = 3.43, p=7.0×10-6<0.001). The longer search time and additional operations did not make them feel burdened with tasks, and they preferred the proposed over the baseline tool. In Table III, the five-grade ratings of all floor plans selected by participants also indicated that they were significantly more satisfied with their selections using our proposed apartment search tool than the baseline tool (M(proposed) = 4.01, M(baseline) = 3.75, p=1.67×10-40<0.001: a t-test was conducted on two related scores; i.e., the average scores of the five floor plans for the five scenarios from all participants using both tools).

## VII. LIMITATIONS

One limitation of our work is that predicted functionality and comfort scores do not come with explanations. People may have different opinions and viewpoints on such subjective scores, and visualizing how the system evaluates floor plan images would be preferred.

Our proposed method has been applied only to apartment floor plans in Japan. As can be seen in the figures, the drawing styles of the floor plan images are very diverse, but they are apparently those in other countries. Therefore, the prediction model need to train for each country. We expect that our segmentation and functionality and comfort prediction models can be used as pre-trained models for fine-tuning, but this is left for our future work.

## VIII. CONCLUSIONS

We created and analyzed a large-scale dataset based on subjective evaluation of real estate floor plan images using crowdsourcing. Our proposed methods for extracting features from graph structures and images of floor plans proved to be effective, as we obtained functionality and comfort prediction models with high accuracy (PCC=0.701). This study is the first work to propose a highly accurate prediction model for dwelling functionality and comfort using machine learning. We applied the results of the prediction model to our new apartment search tool using functionality and comfort as query items, and our user study showed that our tool could provide a better user experience.

## REFERENCES

[1] V. James, S. Wu, A. Gelfand, and C. Sirmans, "Apartment rent prediction using spatial modeling," *Journal of Real Estate Research*, vol. 27, no. 1, pp. 105–136, 2005.
[2] O. Poursaeed, T. Matera, and S. Belongie, "Vision-based real estate price estimation," *Machine Vision and Applications*, vol. 29, no. 4, pp. 667–676, 2018.
[3] M. Heidari, S. Zad, and S. Rafatirad, "Ensemble of supervised and unsupervised learning models to predict a profitable business decision," in *IEEE International IOT, Electronics and Mechatronics Conference (IEMTRONICS)*, 2021, pp. 1–6.
[4] H. Seya and D. Shiroi, "A comparison of residential apartment rent price predictions using a large data set: Kriging versus deep neural network," *Geographical Analysis*, pp. 1–22, 2021.
[5] T. Yamasaki, J. Zhang, and Y. Takada, "Apartment structure estimation using fully convolutional networks and graph model," in *ACM Workshop on Multimedia for Real Estate Tech*, 2018, pp. 1–6.
[6] S. Law, B. Paige, and C. Russell, "Take a look around: Using street view and satellite images to estimate house prices," *ACM Transactions on Intelligent Systems and Technology*, vol. 10, no. 5, Sep. 2019.
[7] M. Zeppelzauer, M. Despotovic, M. Sakeena, D. Koch, and M. Döller, "Automatic prediction of building age from photographs," in *ACM International Conference on Multimedia Retrieval*, 2018, pp. 126–134.
[8] J. H. Bappy, J. R. Barr, N. Srinivasan, and A. K. Roy-Chowdhury, "Real estate image classification," in *IEEE Winter Conference on Applications of Computer Vision (WACV)*, 2017, pp. 373–381.
[9] X. Wang, Y. Takada, Y. Kado, and T. Yamasaki, "Predicting the attractiveness of real-estate images by pairwise comparison using deep learning," in *IEEE International Conference on Multimedia & Expo Workshops (ICMEW)*, 2019, pp. 84–89.
[10] T. Hanazato, Y. Hirano, and M. Sasaki, "Syntactic analysis of large-size condominium units supplied in the tokyo metropolitan area," *Journal of Architecture and Planning*, no. 591, pp. 9–16, 2005.
[11] A. Takizawa, K. Yoshida, and N. Katoh, "Applying graph mining to rent analysis considering room layouts," *Journal of Environmental Engineering (Transaction of AIJ)*, vol. 73, no. 623, pp. 139–146, 2008.
[12] R. Hattori, K. Okamoto, and A. Shibata, "Visualizing the importance of floor-plan image features in rent-prediction models," in *Joint 11th International Conference on Soft Computing and Intelligent Systems and 21st International Symposium on Advanced Intelligent Systems (SCIS-ISIS)*, 2020, pp. 1–3.
[13] R. Hattori, K. Okamoto, and A. Shibata, "Impact analysis of floor-plan images for rent-prediction model (in japanese)," *Journal of Japan Society for Fuzzy Theory and Intelligent Informatics*, vol. 33, no. 2, pp. 640–650, 2021.
[14] K. Solovev and N. Pröllochs, "Integrating floor plans into hedonic models for rent price appraisal," in *Web Conference*, 2021, p. 2838–2847.
[15] G. Csurka, C. Bray, C. Dance, and L. Fan, "Visual categorization with bags of keypoints," *Workshop on Statistical Learning in Computer Vision, ECCV*, pp. 1–22, 2004.
[16] S. Sirmans, D. Macpherson, , and E. Zietz, "The composition of hedonic pricing models," *Journal of Real Estate Literature*, vol. 13, no. 1, p. 1–44, 2005.
[17] Y. Takada, N. Inoue, T. Yamasaki, and K. Aizawa, "Similar floor plan retrieval featuring multi-task learning of layout type classification and room presence prediction," in *IEEE International Conference on Consumer Electronics (ICCE)*, 2018, pp. 1–6.
[18] N. Kato, T. Yamasaki, K. Aizawa, and T. Ohama, "Users' preference prediction of real estates featuring floor plan analysis using floornet," in *ACM Workshop on Multimedia for Real Estate Tech*, 2018, p. 7–11.
[19] N. Kato, T. Yamasaki, K. Aizawa, and T. Ohama, "Users' preference prediction of real estate properties based on floor plan analysis," *IEICE TRANSACTIONS on Information and Systems*, vol. E103-D, no. 2, pp. 398–405, 2020.
[20] M. Yamada, X. Wang, and T. Yamasaki, "Graph structure extraction from floor plan images and its application to similar property retrieval," in *IEEE International Conference on Consumer Electronics*, 2021.
[21] V. Trinh and R. Manduchi, "Semantic interior mapology: A toolbox for indoor scene description from architectural floor plans," *arXiv preprint arXiv:1911.11356*, 2019.
[22] N. Nauata, K.-H. Chang, C.-Y. Cheng, G. Mori, and Y. Furukawa, "House-gan: Relational generative adversarial networks for graph-constrained house layout generation," 2020.
[23] R. Hu, Z. Huang, Y. Tang, O. van Kaick, H. Zhang, and H. Huang, "Graph2plan: Learning floorplan generation from layout graphs," *ACM Transactions on Graphics*, 2020.
[24] S.-T. Yang, F.-E. Wang, C.-H. Peng, P. Wonka, M. Sun, and H.-K. Chu, "Dula-net: A dual-projection network for estimating room layouts from a single rgb panorama," in *IEEE/CVF Computer Vision and Pattern Recognition*, 2019, pp. 3363–3372.
[25] C. Lin, C. Li, and W. Wang, "Floorplan priors for joint camera pose and room layout estimation," 2018.
[26] C. Liu, J. Wu, and Y. Furukawa, "Floornet: A unified framework for floorplan reconstruction from 3d scans," in *European Conference on Computer Vision (ECCV)*, 2018, pp. 203–219.
[27] C. Lin, C. Li, and W. Wang, "Floorplan-jigsaw: Jointly estimating scene layout and aligning partial scans," in *IEEE International Conference on Computer Vision (ICCV)*, 2019, pp. 5674–5683.
[28] J. Chen, C. Liu, J. Wu, and Y. Furukawa, "Floor-sp: Inverse cad for floorplans by sequential room-wise shortest path," in *IEEE International Conference on Computer Vision (ICCV)*, 2019.

[29] Y. Cui, Q. Li, B. Yang, W. Xiao, C. Chen, and Z. Dong, "Automatic 3-d reconstruction of indoor environment with mobile laser scanning point clouds," *IEEE Journal of Selected Topics in Applied Earth Observations and Remote Sensing*, vol. 12, no. 8, pp. 3117–3130, 2019.

[30] A. Phalak, V. Badrinarayanan, and A. Rabinovich, "Scan2plan: Efficient floorplan generation from 3d scans of indoor scenes," *arXiv preprint arXiv:2003.07356*, 2020.

[31] K. Wang, Y.-A. Lin, B. Weissmann, M. Savva, A. X. Chang, and D. Ritchie, "Planit: Planning and instantiating indoor scenes with relation graph and spatial prior networks," *ACM Transactions on Graphics*, vol. 38, no. 4, 2019.

[32] Q. Li, W. Lin, J. Xu, and Y. Fang, "Blind image quality assessment using statistical structural and luminance features," *IEEE Transactions on Multimedia*, vol. 18, no. 12, pp. 2457–2469, 2016.

[33] Q. Wu, H. Li, Z. Wang, F. Meng, B. Luo, W. Li, and K. N. Ngan, "Blind image quality assessment based on rank-order regularized regression," *IEEE Transactions on Multimedia*, vol. 19, no. 11, pp. 2490–2504, 2017.

[34] Z. Fan, T. Jiang, and T. Huang, "Active sampling exploiting reliable informativeness for subjective image quality assessment based on pairwise comparison," *IEEE Transactions on Multimedia*, vol. 19, no. 12, pp. 2720–2735, 2017.

[35] H. Hofbauer, F. Autrusseau, and A. Uhl, "Low quality and recognition of image content," *IEEE Transactions on Multimedia*, 2021.

[36] Y. Li, S. Meng, X. Zhang, M. Wang, S. Wang, Y. Wang, and S. Ma, "User-generated video quality assessment: A subjective and objective study," *IEEE Transactions on Multimedia*, 2021.

[37] Y. Liu, J. Wu, A. Li, L. Li, W. Dong, G. Shi, and W. Lin, "Video quality assessment with serial dependence modeling," *IEEE Transactions on Multimedia*, 2021.

[38] J. Gutierrez, P. Perez, M. Orduna, A. Singla, C. Cortes, P. Mazumdar, I. Viola, K. Brunnstrom, F. Battisti, N. Cieplinska, D. Juszka, L. Janowski, M. I. Leszczuk, A. Adeyemi-Ejeye, Y. Hu, Z. Chen, G. Van Wallendael, P. Lambert, C. Diaz, J. Hedlund, O. Hamsis, S. Fremerey, F. Hofmeyer, A. Raake, P. Cesar, M. Carli, and N. Garcia, "Subjective evaluation of visual quality and simulator sickness of short 360 videos: Itu-t rec. p.919," *IEEE Transactions on Multimedia*, 2021.

[39] P. Lebreton and K. Yamagishi, "Predicting user quitting ratio in adaptive bitrate video streaming," *IEEE Transactions on Multimedia*, vol. 23, pp. 4526–4540, 2021.

[40] Y. Deng, C. C. Loy, and X. Tang, "Image aesthetic assessment: An experimental survey," *IEEE Signal Processing Magazine*, vol. 34, no. 4, pp. 80–106, 2017.

[41] G. Guo, H. Wang, C. Shen, Y. Yan, and H.-Y. M. Liao, "Automatic image cropping for visual aesthetic enhancement using deep neural networks and cascaded regression," *IEEE Transactions on Multimedia*, vol. 20, no. 8, pp. 2073–2085, 2018.

[42] V. Hosu, B. Goldlücke, and D. Saupe, "Effective aesthetics prediction with multi-level spatially pooled features," in *2019 IEEE/CVF Conference on Computer Vision and Pattern Recognition (CVPR)*, 2019, pp. 9367–9375.

[43] P. Lv, J. Fan, X. Nie, W. Dong, X. Jiang, B. Zhou, M. Xu, and C. Xu, "User-guided personalized image aesthetic assessment based on deep reinforcement learning," *IEEE Transactions on Multimedia*, pp. 1–1, 2021.

[44] H. Zhu, Y. Zhou, L. Li, Y. Li, and Y. Guo, "Learning personalized image aesthetics from subjective and objective attributes," *IEEE Transactions on Multimedia*, 2021.

[45] Y. Bai, Z. Zhu, G. Jiang, and H. Sun, "Blind quality assessment of screen content images via macro-micro modeling of tensor domain dictionary," *IEEE Transactions on Multimedia*, vol. 23, pp. 4259–4271, 2021.

[46] B. Pang and L. Lee, "Opinion mining and sentiment analysis," *Found. Trends Inf. Retr.*, vol. 2, no. 1-2, pp. 1–135, Jan. 2008.

[47] X. Lei, X. Qian, and G. Zhao, "Rating prediction based on social sentiment from textual reviews," *IEEE Transactions on Multimedia*, vol. 18, no. 9, pp. 1910–1921, 2016.

[48] F. Chen, R. Ji, J. Su, D. Cao, and Y. Gao, "Predicting microblog sentiments via weakly supervised multimodal deep learning," *IEEE Transactions on Multimedia*, vol. 20, no. 4, pp. 997–1007, 2018.

[49] R. Ji, F. Chen, L. Cao, and Y. Gao, "Cross-modality microblog sentiment prediction via bi-layer multimodal hypergraph learning," *IEEE Transactions on Multimedia*, vol. 21, no. 4, pp. 1062–1075, 2019.

[50] X. Lu, Z. Lin, H. Jin, J. Yang, and J. Z. Wang, "Rapid: Rating pictorial aesthetics using deep learning," in *ACMMM*, 2014, pp. 457–466.

[51] T. Li, B. Ni, M. Xu, M. Wang, Q. Gao, and S. Yan, "Data-driven affective filtering for images and videos," *IEEE TCYB*, 2015 (in press).

[52] J. Yang, D. She, M. Sun, M.-M. Cheng, P. L. Rosin, and L. Wang, "Visual sentiment prediction based on automatic discovery of affective regions," *IEEE Transactions on Multimedia*, vol. 20, no. 9, pp. 2513–2525, 2018.

[53] X. Yao, D. She, H. Zhang, J. Yang, M.-M. Cheng, and L. Wang, "Adaptive deep metric learning for affective image retrieval and classification," *IEEE Transactions on Multimedia*, vol. 23, pp. 1640–1653, 2021.

[54] J. Yang, D. She, M. Sun, M.-M. Cheng, P. L. Rosin, and L. Wang, "Visual sentiment prediction based on automatic discovery of affective regions," *IEEE Transactions on Multimedia*, vol. 20, no. 9, pp. 2513–2525, 2018.

[55] S. Ruan, K. Zhang, L. Wu, T. Xu, Q. Liu, and E. Chen, "Color enhanced cross correlation net for image sentiment analysis," *IEEE Transactions on Multimedia*, pp. 1–1, 2021.

[56] Y. Su, W. Zhao, P. Jing, and L. Nie, "Exploiting low-rank latent gaussian graphical model estimation for visual sentiment distribution," *IEEE Transactions on Multimedia*, pp. 1–1, 2022.

[57] H. Zhang and M. Xu, "Multiscale emotion representation learning for affective image recognition," *IEEE Transactions on Multimedia*, pp. 1–1, 2022.

[58] S. Zhang, S. Zhang, T. Huang, and W. Gao, "Speech emotion recognition using deep convolutional neural network and discriminant temporal pyramid matching," *IEEE Transactions on Multimedia*, vol. 20, no. 6, pp. 1576–1590, 2018.

[59] T. Li and M. Ogihara, "Toward intelligent music information retrieval," *IEEE TMM*, vol. 8, no. 3, pp. 564–574, June 2006.

[60] G. Irie, T. Satou, A. Kojima, T. Yamasaki, and K. Aizawa, "Affective audio-visual words and latent topic driving model for realizing movie affective scene classification," *IEEE TMM*, vol. 12, no. 6, pp. 523–535, Oct 2010.

[61] R. Teixeira, T. Yamasaki, and K. Aizawa, "Affective determination of video clips by low level audiovisual features -a dimensional and categorial approach-," *Multimedia Tools and Applications*, 2011.

[62] M. Soleymani, M. Pantic, and T. Pun, "Multimodal emotion recognition in response to videos," *IEEE TAC*, vol. 3, no. 2, pp. 211–223, 2012.

[63] T. Liu, J. Wan, X. Dai, F. Liu, Q. You, and J. Luo, "Sentiment recognition for short annotated gifs using visual-textual fusion," *IEEE Transactions on Multimedia*, vol. 22, no. 4, pp. 1098–1110, 2020.

[64] Q. Fang, C. Xu, J. Sang, M. S. Hossain, and G. Muhammad, "Word-of-mouth understanding: Entity-centric multimodal aspect-opinion mining in social media," *IEEE Transactions on Multimedia*, vol. 17, no. 12, pp. 2281–2296, 2015.

[65] W. Guo, Y. Zhang, X. Cai, L. Meng, J. Yang, and X. Yuan, "Ld-man: Layout-driven multimodal attention network for online news sentiment recognition," *IEEE Transactions on Multimedia*, vol. 23, pp. 1785–1798, 2021.

[66] W. Nie, R. Chang, M. Ren, Y. Su, and A. Liu, "I-gcn: Incremental graph convolution network for conversation emotion detection," *IEEE Transactions on Multimedia*, pp. 1–1, 2021.

[67] P. Parmar and B. T. Morris, "What and how well you performed? a multitask learning approach to action quality assessment," in *2019 IEEE/CVF Conference on Computer Vision and Pattern Recognition (CVPR)*, 2019, pp. 304–313.

[68] D. Liu, Q. Li, T. Jiang, Y. Wang, R. Miao, F. Shan, and Z. Li, "Towards unified surgical skill assessment," in *2021 IEEE/CVF Conference on Computer Vision and Pattern Recognition (CVPR)*, 2021, pp. 9517–9526.

[69] M. Redi, N. O'Hare, R. Schifanella, M. Trevisiol, and A. Jaimes, "6 seconds of sound and vision: Creativity in micro-videos," in *2014 IEEE Conference on Computer Vision and Pattern Recognition*, 2014, pp. 4272–4279.

[70] B. Xia, X. Wang, T. Yamasaki, K. Aizawa, and H. Seshime, "Deep neural network-based click-through rate prediction using multimodal features of online banners," in *2019 IEEE Fifth International Conference on Multimedia Big Data (BigMM)*, 2019, pp. 162–170.

[71] J. Ikeda, H. Seshime, X. Wang, and T. Yamasaki, "25th international conference on pattern recognition (icpr)," in *25th International Conference on Pattern Recognition (ICPR)*, 2995-3002, pp. 2995–3002.

[72] S. Oyama and T. Yamasaki, "Visual clarity analysis and improvement support for presentation slides," in *2019 IEEE Fifth International Conference on Multimedia Big Data (BigMM)*, 2019, pp. 421–428.

[73] S. Yi, J. Matsugami, and T. Yamasaki, "Assessment system of presentation slide design using visual and structural features," *IEICE Transactions on Information and Systems*, vol. E105-D, no. 3, 2022.

[74] C. Liu, J. Wu, P. Kohli, and Y. Furukawa, "Raster-to-vector: Revisiting floorplan transformation," in *2017 IEEE International Conference on Computer Vision (ICCV)*, 2017, pp. 2214–2222.

[75] Z. Zeng, X. Li, Y. K. Yu, and C.-W. Fu, "Deep floor plan recognition using a multi-task network with room-boundary-guided attention," in *IEEE/CVF International Conference on Computer Vision (ICCV)*, 2019.

[76] C. Mura, R. Pajarola, K. Schindler, and N. Mitra, "Walk2map: Extracting floor plans from indoor walk trajectories," in *Computer Graphics Forum*, vol. 40, no. 2, 2021, pp. 375–388.

[77] L.-C. Chen, G. Papandreou, I. Kokkinos, K. Murphy, and A. L. Yuille, "Deeplab: Semantic image segmentation with deep convolutional nets, atrous convolution, and fully connected crfs," *IEEE Transactions on Pattern Analysis and Machine Intelligence*, vol. 40, no. 4, pp. 834–848, 2017.

[78] X. Yan and J. Han, "gspan: Graph-based substructure pattern mining," in *IEEE International Conference on Data Mining*, 2002, pp. 721–724.

[79] G. Dong and J. Li, "Efficient mining of emerging patterns: Discovering trends and differences," in *fifth ACM SIGKDD international conference on Knowledge discovery and data mining*, 1999, pp. 43–52.

[80] K. Ohara, T. Yamasaki, and K. Aizawa, "An intuitive system for searching apartments using floor plans and areas of rooms," *78th national convention of IPSJ*, vol. 2016, no. 1, pp. 311–312, mar 2016.

[81] J. J. McGregor, "Backtrack search algorithms and the maximal common subgraph problem," *Software: Practice and Experience*, vol. 12, no. 1, pp. 23–34, 1982.

[82] K. He, X. Zhang, S. Ren, and J. Sun, "Deep residual learning for image recognition," in *IEEE conference on computer vision and pattern recognition (CVPR)*, 2016, pp. 770–778.

[83] A. Krizhevsky, I. Sutskever, and G. E. Hinton, "Imagenet classification with deep convolutional neural networks," in *Advances in neural information processing systems*, 2012, pp. 1097–1105.

[84] S. Ioffe and C. Szegedy, "Batch normalization: Accelerating deep network training by reducing internal covariate shift," *arXiv preprint arXiv:1502.03167*, 2015.

[85] N. Srivastava, G. Hinton, A. Krizhevsky, I. Sutskever, and R. Salakhutdinov, "Dropout: a simple way to prevent neural networks from overfitting," *The journal of machine learning research*, vol. 15, no. 1, pp. 1929–1958, 2014.